\documentstyle[array,epsfig,twocolumn,pra,aps]{revtex}

\begin{document}
\draft
\wideabs{
  \title{
Quantum nonlocality test for continuous-variable
states with dichotomic observables}
  \author{H. Jeong$^1$, W. Son$^1$, M. S. Kim$^1$, D. Ahn$^2$, and
  {\v C}. Brukner$^3$}  
  \address{$^1$School of Mathematics and Physics, Queen's University,
    Belfast BT7 1NN, United Kingdom \\
$^2$Institute of Quantum Information Processing and Systems,
University of Seoul, Seoul, 130-743, Korea
\\
$^3$Instutut f${\ddot u}$r Experimentalphysik, Universit${\ddot a}$t
Wien, Boltzmangasse 5, A-1090, Austria}
  \date{\today}  
  \maketitle
\begin{abstract}
  There have been theoretical and experimental studies on quantum
  nonlocality for continuous variables, based on dichotomic
  observables.  In particular, we are interested in two cases of
  dichotomic observables for the light field of continuous variables:
  One case is even and odd numbers of photons and the other case is no
  photon and the presence of photons.  We analyze various observables
  to give the maximum violation of Bell's inequalities for
  continuous-variable states.  We discuss an observable which gives
  the violation of Bell's inequality for any entangled pure continuous
  variable state.  However, it does not have to be a maximally
  entangled state to give the maximal violation of the Bell's
  inequality.  This is attributed to a generic problem of testing the
  quantum nonlocality of an infinite-dimensional state using a
  dichotomic observable.
\end{abstract}


}

\section{introduction}
The paradox suggested by Einstein, Podolsky and Rosen aroused
controversy about nonlocality of quantum states \cite{EPR}.  Bell
proposed a remarkable inequality imposed by a local hidden variable
theory \cite{Bell}, which enables a quantitative test on quantum
nonlocality.  Numerous theoretical studies and experimental
demonstrations have been performed to understand nonlocal properties
of quantum states.  Various versions of Bell's inequality
\cite{CHSH,CH} followed the original one \cite{Bell}.

Gisin and Peres  found pairs of observables whose
correlations violate Bell's inequality for
a discrete $N$-dimensional entangled state \cite{GP}.
Banaszek and W{\' o}dkiewicz (BW) studied Bell's inequality for
continuous-variable states, in terms of Wigner representation in phase
space based upon parity measurement and displacement operation
\cite{BW}.  This is useful because of its experimental 
relevance but does
not lead to maximal violation for the original Einstein-Podolsky-Rosen
(EPR) state \cite{Kuzmich}. Recently, Chen {\it et al.} studied Bell's inequality of
continuous-variable states  \cite{Chen01} using their newly defined Bell operator \cite{Chen01,halv}.
In contrast to the operators in BW formalism, the pseudo-spin operators are
not experimentally easy to realize but the EPR state
 can
maximally violate Bell's inequality in their framework \cite{Chen01}.

In this paper, we relate Chen {\it et al.}'s ``pseudo-spin'' Bell
operator to one of Gisin and Peres for a finite-dimensional state to
bridge the gap between the discussions for the nonlocality of finite
and infinite dimensional (or continuous-variable) systems.  The origin
of the pseudospin operator is attributed to the limiting case of
Gisin-Peres observable \cite{GP}.  
We investigate various versions of Clauser, Horne,
Shimony and Holt 
(CHSH)'s inequality for continuous-variable states.
 It is pointed out that the BW formalism
can be generalized to obtain a larger Bell violation \cite{Derek}, but
it cannot give the maximal violation for the EPR state even
in the generalized version.  We analyze the reason why the EPR state
cannot maximally violate Bell's inequality in the generalized BW
formalism.  We compare the EPR state with an entangled state of two
coherent states \cite{Sanders}.  
 In contrast to the EPR state, the entangled
coherent state  shows
the maximal Bell violation for certain limit  both for the generalized BW and Chen
{\it et al.}'s formalism.
  We also
investigate Clauser and Horne (CH)'s version of Bell's inequality.  We
find the upper and lower bounds for the Bell-CH inequality and test
whether 
the values for continuous-variable states reach these bounds.

\section{origin of pseudospin operator}
Chen {\it et al.} introduced a pseudospin operator ${\bf
  s}=(s_x,s_y,s_z)$ for a nonlocality test of continuous variables as a
direct analogy of a spin-1/2 system \cite{Chen01,halv}:
\begin{eqnarray}
\label{eq:A}
&&s_z=\sum_{n=0}^\infty\big(|2n+1\rangle\langle 2n+1|-|2n\rangle\langle 2n|\big),\\
&&s_x\pm s_y=2s_\pm,\label{eq:sxy}\\
&&{\bf a}\cdot {\bf s}=s_z \cos\theta+\sin\theta(e^{i \varphi}s_-+e^{-i \varphi}s_+),
\label{eq:as}
\end{eqnarray}
where $s_-=\sum_{n=0}^\infty|2n\rangle\langle 2n+1|=(s_+)^\dagger$ and
${\bf a}$ is a unit vector. 
The Bell-CHSH operator 
 based upon the pseudospin operator is then defined as  \cite{CHSH,Chen01}
\begin{eqnarray}
&&{\cal B}=({\bf a\cdot s}_1)\otimes({\bf b\cdot s}_2)
+({\bf a\cdot s}_1)\otimes({\bf b^\prime\cdot s}_2)\nonumber\\
&&~~~~~~+({\bf a^\prime\cdot s}_1)\otimes({\bf b\cdot s}_2)
-({\bf a^\prime\cdot s}_1)\otimes({\bf b^\prime\cdot s}_2),
\end{eqnarray}
where 1 and 2 denote two different modes and ${\bf
  a}^\prime$, ${\bf b}$ and ${\bf b}^\prime$ are unit vectors.

Bell's inequality imposed by local hidden variable theory is then
$|\langle {\cal B}\rangle|\leq2$.  In this formalism, the violation of
the inequality is limited by Cirel'son bound $|\langle {\cal
  B}\rangle|\leq2\sqrt{2}$ \cite{Chen01,C80}.  
It was found that a two-mode squeezed state
\begin{equation}
|{\rm TMSS}\rangle=\sum_{n=0}^{\infty}\frac{(\tanh r)^n}{\cosh r}|n\rangle|n\rangle, \label{eq:nopa}
\end{equation}
where $|n\rangle$ is a number state and $r$ is the squeezing
parameter, maximally violates Bell's inequality, {\it i.e.} $|\langle {\cal B}\rangle|_{max}\rightarrow 2\sqrt{2}$  when 
$r$ becomes infinity \cite{Chen01}.  Note that the two-mode squeezed state (\ref{eq:nopa})
becomes the original EPR state when $r\rightarrow\infty$
\cite{Braunstein98}. 

 Gisin and Peres found pairs of observables whose
correlations violate Bell's inequality for
an $N$-dimensional entangled state \cite{GP} 
\begin{equation} 
|\Psi\rangle=\sum_{n=0}^{N-1} c_n|\phi_n\rangle|\psi_n\rangle,
\end{equation}
 where $\{|\phi_n\rangle\}$ and
$\{|\psi_n\rangle\}$ are any orthonormal bases.  Further they showed that the violation of Bell's
inequality is maximal in the case of a spin-$j$ singlet state for an
even $j$.
The Gisin-Peres observable is 
\begin{equation}
\label{GP-observable}
A(\theta)=\Gamma_x\sin\theta+\Gamma_z\cos\theta+{\cal E},\\
\end{equation}
where $\Gamma_x$ and $\Gamma_z$ are block-diagonal matrices in which
each block is an ordinary Pauli matrix, $\sigma_x$ and $\sigma_z$,
respectively.
 ${\cal E}$ is a matrix whose only non-vanishing element
is ${\cal E}_{N-1,N-1}=1$ when $N$ is odd and ${\cal E}$ is zero when
$N$ is even. 
 The Bell operator is then defined as 
\begin{eqnarray}
\label{Bell-GP}
&&{\cal B}_{GP}=({\bf a}\cdot A_1)\otimes({\bf b}\cdot A_2)
+({\bf a}\cdot A_1)\otimes({\bf b}^\prime\cdot A_2)\nonumber\\
&&~~~~~~+({\bf a}^\prime\cdot A_1)\otimes({\bf b}\cdot A_2)
-({\bf a}^\prime\cdot A_1)\otimes({\bf b}^\prime\cdot A_2),
\end{eqnarray}
where $A$ represents the Gisin-Peres observable $A(\theta)$.
It was Gisin \cite{Gisin} who showed any entangled pure state violates a Bell's inequality.
Later, Gisin and Peres \cite{GP} found the observable
(\ref{GP-observable}) to give
the violation of Bell's inequality for any $N$-dimensional entangled pure state.

In the limit $N\rightarrow\infty$, we find that
$\Gamma_x$ and $\Gamma_z$ become pseudospin operators $s_x$ and $s_z$
in Eq.~(\ref{eq:sxy})
and $A(\theta)$ becomes ${\bf a}\cdot{\bf s}$ (with $\varphi=0$) in Eq.~(\ref{eq:as}).
Note that the effect of ${\cal E}$
vanishes for $N\rightarrow\infty$.
Understanding Chen {\it et al.}'s observables as a limiting case of Gisin-Peres observable defined
for a finite discrete system, it is now straightforward to show that the EPR state
maximally violates Bell's inequality as the EPR state $\sum_{n=0}^{\infty}|n\rangle|n\rangle$
is the {\em infinite-dimensional singlet state}.  
Extending the Gisin and Peres' argument, we can make a 
remark: Any bipartite pure infinite-dimensional entangled state violates Bell's inequality for observables
based on the pseudospin observables.  

\section{The Bell-CHSH inequalities for continuous variables}

\subsection{The two-mode squeezed state}

 Banaszek and W{\' o}dkiewicz  studied Bell's inequality for
 continuous-variable systems based upon parity measurement and displacement
operation \cite{BW}:
\begin{eqnarray}
&&\Pi(\alpha)=\Pi^+(\alpha)-\Pi^-(\alpha) \nonumber\\
&&~~~~~~~=D(\alpha)\sum_{n=0}^\infty\Big(|2n\rangle\langle 2n|
  -|2n+1\rangle\langle 2n+1|\Big)D^\dagger(\alpha)
\label{eq:bw}
\end{eqnarray}
where $D(\alpha)$ is the displacement operator $D(\alpha)=
\exp[\alpha \hat{a}^\dagger-\alpha^* \hat{a}]$ for bosonic operators $\hat{a}$ and 
$\hat{a}^\dag$. 
It should be pointed out that in order to maximize the violation of
Bell's inequality, 
 the BW 
formalism needs to be generalized to write the Bell operator as
\cite{Derek}
\begin{eqnarray}
&&{\cal B}_{BW}=\Pi_1(\alpha)\Pi_2(\beta)+\Pi_1(\alpha^\prime)\Pi_2(\beta)
\nonumber\\
&&~~~~~~~~~~~~~~+\Pi_1(\alpha)\Pi_2(\beta^\prime)-
\Pi_1(\alpha^\prime)\Pi_2(\beta^\prime).
\end{eqnarray}
 BW 
assumed two of the four parameters equal to zero as
$\alpha=\beta=0$.  
The Bell-CHSH inequality can then be represented by the Wigner function as
\begin{eqnarray}
&&|\langle{\cal
  B}_{BW}\rangle|=\frac{\pi^2}{4}|W(\alpha,\beta)+W(\alpha,\beta^\prime)
\nonumber\\
&&~~~~~~~~~~~~~~~~~~+W(\alpha^\prime,\beta)-W(\alpha^\prime,\beta^\prime)|\leq2,
\label{eq:abft}
\end{eqnarray}
where $W(\alpha,\beta)$ represents the Wigner function of a given
state.  Using $\Pi_1(\alpha)\Pi_1(\alpha)=\Pi_2(\alpha)\Pi_2(\alpha)=\openone$,
it is straightforward to check the Cirel'son bound $|\langle{\cal B}_{BW}\rangle|\leq
2\sqrt{2}$ 
in the generalized BW 
formalism.

\begin{figure}
\centerline{\scalebox{.57}{
\includegraphics{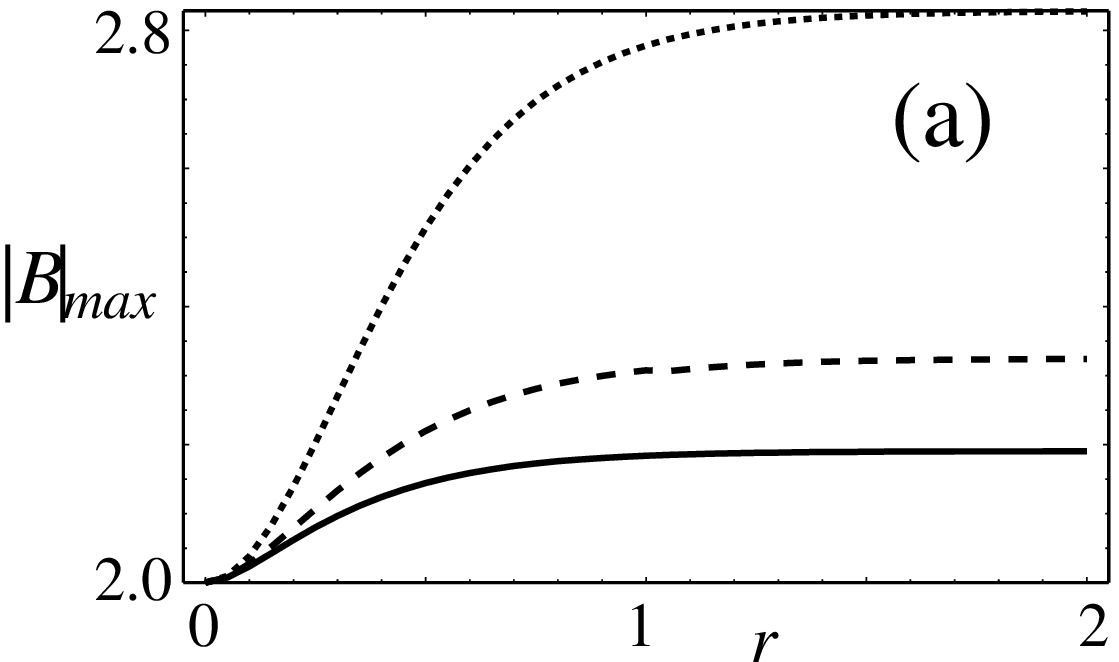}
}} \vspace{0.7cm}
\centerline{\scalebox{.6}{
\includegraphics{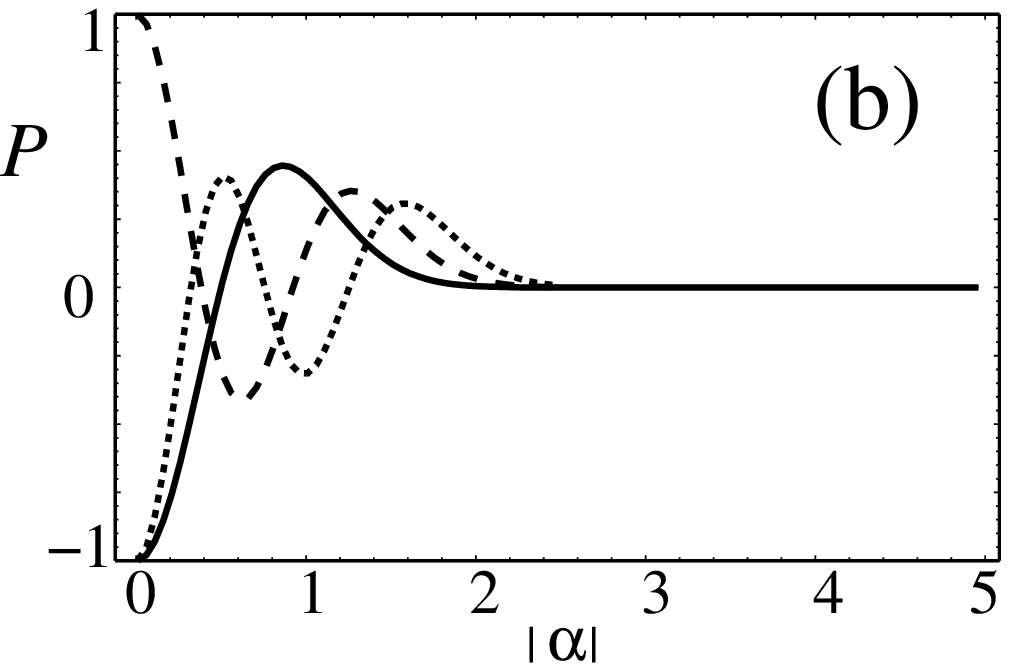}
}} \vspace{0.2cm}
\caption{(a) 
  The maximized value of absolute Bell function $|B|_{max}$ for a
  two-mode squeezed state vs the squeezing parameter $r$ in the BW
  (solid line), the generalized BW (dashed), and Chen {\it et al.}'s
  (dotted) formalisms.  It is shown that the EPR state does
  not maximally violate Bell's inequality in the generalized BW
  formalism.  (b) The expectation value $P$ of BW's observable for
  number states of $n=1$ (solid), $n=2$ (dashed), and $n=3$ (dotted)
  is plotted against the absolute displacement parameter $|\alpha|$.  }\label{fig1}
\end{figure}

The Wigner function of the two-mode squeezed state is \cite{BN}
\begin{eqnarray}
&&W_{TMSS}(\alpha,\beta)=\frac{4}{\pi^2}\exp[-2\cosh 2r(|\alpha|^2
+|\beta|^2)\nonumber\\
&&~~~~~~~~~~~~~~~~~~~~~~~~~~~~~~~~+2\sinh 2r 
(\alpha\beta+\alpha^*\beta^*)],
\end{eqnarray}
with which the Bell function
$B_{BW}\equiv\langle{\cal B}_{BW}\rangle$ can be calculated.
 In the infinite squeezing limit,
 the absolute Bell function maximizes as $|B_{BW}|_{max}\rightarrow8/3^{9/8}\simeq 2.32$ at
 $\alpha=-\alpha^\prime=\beta^\prime/2=\sqrt{(\ln 3)/16\cosh 2r}$ and
 $\beta=0$.  This shows that the EPR state does not maximally violate
 Bell's inequality in the generalized BW formalism.  In Fig.~1(a),
 using the generalized BW formalism, the maximized value
$|B_{BW}|_{max}$ is plotted
 for the two-mode squeezed state
 and compared with the violation of Bell's inequality based on 
other formalisms.
(The method of steepest descent \cite{nume} 
is used 
in Fig.~\ref{fig1}(a) and other figures in the paper to get 
the maximized value of violation within the formalism.)

The reason why the generalized BW formalism does not give the maximum
violation for the EPR state can be explained as follows.  The operator
$s_z$ in Eq.~(\ref{eq:A}) is equivalent to BW's observable
$\Pi(\alpha)$ when $\alpha=0$ except a trivial sign change. The main
difference is that BW use the displacement operator while Chen {\it et
  al.} use the direct analogy of the rotation of spin operators.  When
the Gisin-Peres observable $A(\theta)$ (or equivalently pseudospin
observable ${\bf a}\cdot {\bf s}$ with $\varphi=0$) is applied on
an arbitrary state $\sum_{n=0}^\infty f(n)|n\rangle$, where $f(n)$ is an arbitrary
function, we obtain
\begin{eqnarray}
&&A(\theta)\sum_{n=0}^\infty
f(n)|n\rangle=\sqrt{2}\cos(\theta-\pi/4)\sum_{n=0}^\infty 
f(2n)|2n\rangle\nonumber\\
&&~~~~~~~~~~~~~
  +\sqrt{2}\sin(\theta-\pi/4)\sum_{n=0}^\infty 
f(2n+1)|2n+1\rangle.
\label{eq:sd}
\end{eqnarray}
The operator $A(\theta)$ rotates
$\sum f(n)|n\rangle$ into even and odd parity states; the pseudospin observable
(\ref{eq:as}) can completely flip the parity of any given state by
changing the angle. 
 Note that the only
measurement applied to the nonlocality test here is the parity
measurement.  Differently from the pseudospin operator, BW's
observable $\Pi(\alpha)$ does not assure the complete parity change,
which makes it impossible to find the maximal Bell violation of the
two-mode squeezed state.  In the two-mode squeezed state, orthogonal
number states, which have well defined parity, are the entangled
elements.  The expectation value of BW's observable for a number
state is obtained as \cite{OR}
\begin{eqnarray}
&&P(n,|\alpha|)=\langle n|\Pi(\alpha)|n\rangle \nonumber\\
&&~~~~~~~=\frac{e^{-|\alpha|^2}|\alpha|^{2n}}{n!}\sum_{k=0}^\infty\Big\{\frac{(2k)!}{|\alpha|^{4k}}\big(L_{2k}^{(n-2k)}(|\alpha|^2)\big)^2
\nonumber\\
&&~~~~~~~~~~~~~~~~~~~~~-\frac{(2k+1)!}{|\alpha|^{4k+2}}\big(L_{2k+2}^{(n-2k-1)}
(|\alpha|^2)\big)^2\Big\},
\end{eqnarray}
where $L_q^{(p)}(x)$ is an associated Laguerre polynomial.
We numerically assess $P(n,|\alpha|)$ for some different numbers  and check that the parity of
the number states cannot be perfectly flipped by changing the
parameter $\alpha$ of the displacement
operator $D(\alpha)$ as shown in Fig.~1(b).

\subsection{The entangled coherent state}
\begin{figure}
\centerline{
{\scalebox{.57}
{\includegraphics{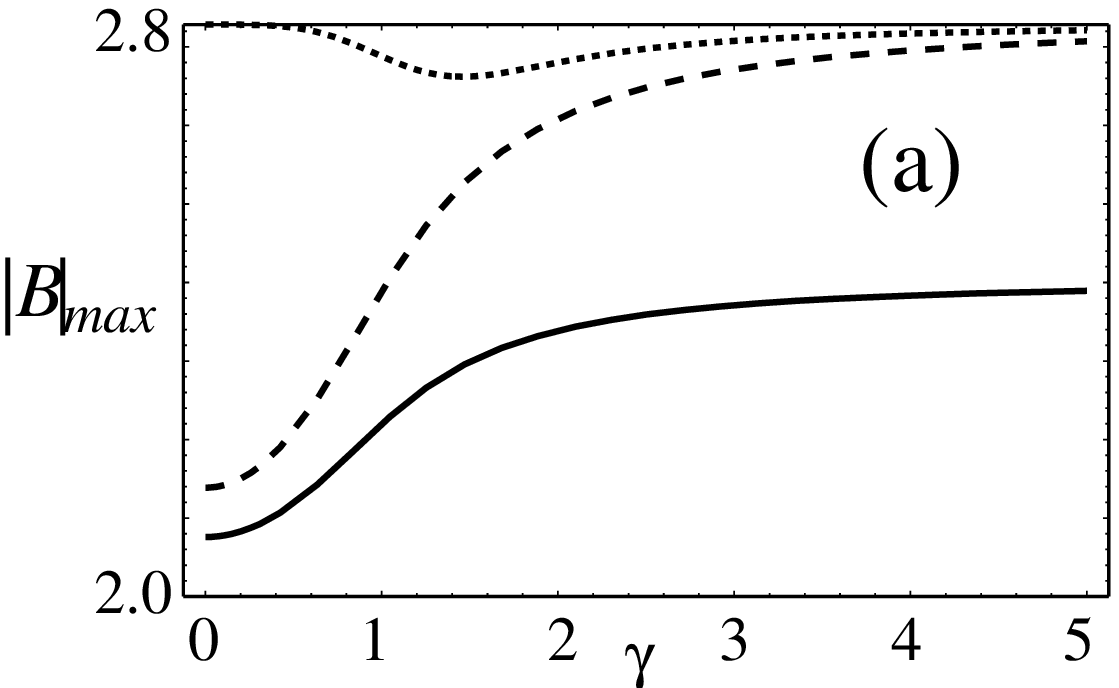}
}}}\vspace{0.7cm}
\centerline{
{\scalebox{.6}
{\includegraphics{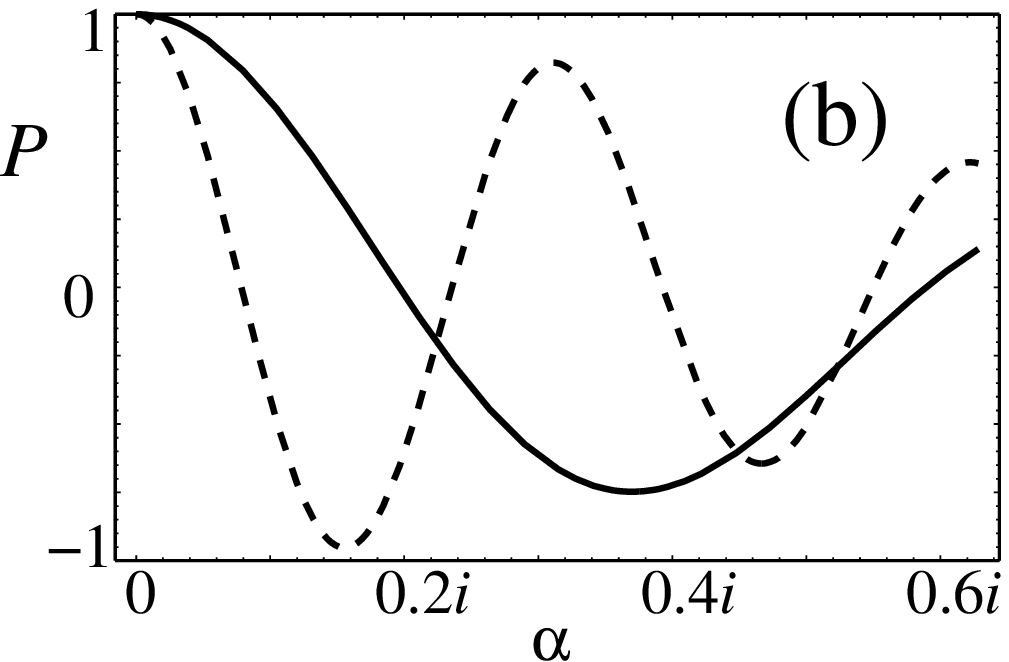}
}}}\vspace{0.2cm}
\caption{
  (a) The maximized value of absolute Bell function $|B|_{max}$ for an
  entangled coherent state is plotted against its coherent amplitude $\gamma$ using the
  BW (solid), the generalized BW (dashed), and Chen {\it et al.}'s
  (dotted) formalisms.  The entangled coherent state maximally violates
  Bell's inequality in the generalized BW formalism for
  $\gamma\rightarrow\infty$ and in the Chen {\it et al.}'s formalism
  both for $\gamma\rightarrow0$ (but $\gamma\neq 0$) and for $\gamma\rightarrow\infty$. (b)
  The expectation value $P$ of BW's observable for the even cat
  state is plotted against $\alpha$ for $\gamma=2$ (solid) and $\gamma=5$ (dashed).
  For $\gamma\gg1$, the displacement operator acts like a
  rotation so that the parity of the even and odd cat states may be
  well flipped. }
\label{fig2}
\end{figure}

The entangled coherent state \cite{Sanders}
 is another important continuous-variable entangled
state.  Many possible applications to quantum information processing have been studied
utilizing entangled coherent states
 \cite{t1}.
The entangled coherent state $|{\rm ECS}\rangle$ can be defined as
\begin{eqnarray}
|{\rm ECS}\rangle&=&{\cal N}(|\gamma\rangle|-\gamma\rangle
 -|-\gamma\rangle|\gamma\rangle), \label{ecs1}\\
|\gamma\rangle&=&
e^{-|\gamma|^2/2}\sum_{n=0}^{\infty}\frac{\gamma^n}{\sqrt{n!}}|n\rangle,
\end{eqnarray}
where ${\cal N}$ is a normalization factor and $|\gamma\rangle$ is a
coherent state with $\gamma\neq0$.
For the case of the entangled coherent state,
the Bell function  in the generalized BW formalism (\ref{eq:abft}) can be calculated from its Wigner function:
\begin{eqnarray}
&&W_{ECS}(\alpha,\beta)=4{\cal N}^2\Big\{
\exp[-2|\alpha-\gamma|^2-2|\beta+\gamma|^2]\nonumber\\
&&~~~~~~~~~~~~~~~~~~~~~~~~~~+\exp[-2|\alpha+\gamma|^2-2|\beta-\gamma|^2]\nonumber\\
&&~~-\exp[-2(\alpha-\gamma)(\alpha^*+\gamma)-2(\beta+\gamma)(\beta^*-\gamma)-4\gamma^2]\nonumber\\
&&~~-\exp[-2(\alpha^*-\gamma)(\alpha+\gamma)-2(\beta^*+\gamma)(\beta-\gamma)-4\gamma^2]
\Big\},\nonumber\\
\end{eqnarray}
where $\gamma$ is assumed to be real for simplicity.
We find that the Bell function
approaches to
$2\sqrt{2}$ for $\gamma\rightarrow\infty$  \cite{Derek}  at 
$\alpha=0$, $\beta=5\pi/16\gamma$, $\alpha^\prime=\pi/8\gamma$ and
$\beta^\prime=3\pi/16\gamma$ as shown in Fig.~\ref{fig2}(a).

The entangled coherent state can be represented in the
 $2\times2$-Hilbert space as
\begin{equation}
|{\rm ECS}\rangle=\frac{1}{\sqrt{2}}(|e\rangle|d\rangle
 -|d\rangle|e\rangle),
\label{ecs2}
\end{equation}
where  $|e\rangle={\cal N_+}(|\gamma\rangle+|-\gamma\rangle)$ and
 $|d\rangle={\cal N_-}(|\gamma\rangle-|-\gamma\rangle)$ are even and
 odd cat states with  normalization factors  ${\cal N_+}$ and ${\cal N_-}$.
Note that these states 
 form an orthogonal basis, regardless of the value of $\gamma$,
which span the two-dimensional Hilbert space.
Suppose that an ideal rotation $R_x(\theta)$ around the $x$ axis,
\begin{eqnarray}
  R_x(\theta)|e\rangle&=&\cos\theta|e\rangle+i\sin\theta|d\rangle,
  \nonumber \\ 
\label{eq:uo}
R_x(\theta)|d\rangle&=&i\sin\theta|e\rangle+\cos\theta|d\rangle,
\end{eqnarray}
can be performed on the both sides of the entangled coherent state
(\ref{ecs2}).  Because the state (\ref{ecs2}) is the same as the
EPR-Bohm state of a two-qubit system, it can be easily proved that it
maximally violates the Bell's inequality, {\em i.e.}, 
the maximized Bell function is $2\sqrt{2}$.  Remarkably, it is known that the
displacement operator acts like the rotation $R_x(\theta)$ on the even
and odd cat states for $\gamma\gg1$ \cite{Derek,c1}.  The fidelity can
be checked that $|\langle
e|D^\dagger(i\alpha_i)R_x(\theta)|e\rangle|^2 =|\langle
d|D^\dagger(i\alpha_i)R_x(\theta)|d\rangle|^2 \rightarrow1$ for
$\gamma\rightarrow\infty$, where $\theta=2\gamma\alpha_i$
and $\alpha_i$ is real.  As a result, the parity
of the even and odd cat states, which are the orthogonal entangled
elements in the entangled coherent state, can be perfectly flipped by
the displacement operator for $\gamma\rightarrow\infty$ as is implied
in Fig.~2(b) \cite{Derek}.  This property enables the maximal Bell
violation of the entangled coherent state for a large coherent
amplitude.

In the pseudospin formalism, the correlation function $E(\theta_1,\varphi_1,\theta_2,\varphi_2)=\langle {\rm
  ECS}|s_1(\theta_1,\varphi_1)
\otimes s_2(\theta_2,\varphi_2)|{\rm ECS}\rangle$ of the entangled
coherent state is 
\begin{eqnarray}
&&E(\theta_1,\varphi_1,\theta_2,\varphi_2)=-\cos\theta_1\cos\theta_2
\nonumber\\
&&~~~~~~~~~~~~~~~~~~~~~~~~-K(\gamma)\cos(\varphi_1-\varphi_2)\sin\theta_1\sin\theta_2,\nonumber\\
&&~~~~~K(\gamma)=\frac{\cosh \gamma^2\sinh \gamma^2}
{(\sum_{n=0}^\infty\frac{\gamma^{4n+1}}{\sqrt{(2n)!(2n+1)!}})^2},
\label{E&K}
\end{eqnarray}
where  $0< K(\gamma)<1$, and $K(\gamma)$ approaches to 1 when $\gamma\rightarrow0$ 
(but $\gamma\neq 0$) and
$\gamma\rightarrow\infty$.
The maximized value of the Bell function $B=\langle {\cal B}\rangle$ is obtained from Eq.~(\ref{E&K}) as
\begin{equation}
|B|_{max} 
=2\sqrt{1+K(\gamma)^2}
\end{equation}
by setting $\theta_1=0$, $\theta^\prime_1=\pi/2$,
$\theta_2=-\theta_2^\prime$ and $\varphi_1=\varphi_2=0$.  Then, the
maximal violation is found for the two extreme cases,
$\gamma\rightarrow0$ and $\gamma\rightarrow\infty$. 
When $\gamma$ is small, the entangled coherent state is not
maximally entangled in an infinite-dimensional Hilbert space as
tracing the state over one mode variables the von Neumann entropy is
not infinite.
It is interesting to note that the non-maximally
entangled state maximally violates Bell's inequality.
  We attribute
this mismatch to the dichotomic nature of the test of quantum nonlocality for
an infinite-dimensional system.
However, the entangled coherent state
is maximally entangled in the $2\times2$ Hilbert space
but it does not always maximally violate the Bell-CHSH inequality as shown in
Fig.~2(a).  This shows that the pseudospin formalism is not a
`perfect' analogy of a two-qubit system when a qubit is composed of
two orthogonal even and odd cat states.  The pseudospin operator ${\bf
  a\cdot s}$ (with $\varphi=0$) in Eq.~(\ref{eq:as}) can be written as
${\bf a\cdot s}=U(\theta)s_z$ where a unitary rotation $U(\theta)$ is
\begin{eqnarray}
&&U(\theta)|2n+1\rangle=\cos\theta|2n+1\rangle+\sin\theta|2n\rangle, \label{RR1}\\
&&U(\theta)|2n\rangle=-\sin\theta|2n+1\rangle+\cos\theta|2n\rangle.
\label{RR2}
\end{eqnarray}
The even (odd) cat state does not flip into the odd (even) cat state
by $U(\theta)$; it is only the parity of the given state which
changes.  The fidelity between the `rotated' odd cat state and the
even cat state is obtained as
\begin{equation}
|\langle d| U(\pi/2)|e\rangle|^2=K(\gamma)
\label{redby}
\end{equation}
which is smaller than 1.  It is clear that $|e\rangle$ and $|d\rangle$
are well flipped to each other only for the limiting cases of
$\gamma\rightarrow0$ and $\gamma\rightarrow\infty$.  In other word,
the rotation may get the given states out of the $2\times2$ space
spanned by $|e\rangle$ and $|d\rangle$. Note, for example, that
$U(\pi/2)|e\rangle$ cannot be represented by a linear superposition of
$|e\rangle$ and $|d\rangle$.

\section{The Clauser-Horne inequality} 
We have studied quantum nonlocality of
continuous-variable states using the Bell-CHSH inequality \cite{CHSH}
and all the arguments have been based upon the parity measurement. The
Clauser and Horne's version of the Bell's inequality \cite{CH} can
also be considered to test the nonlocality of continuous-variable
states with photon number measurement \cite{BW}. We will investigate
the Bell-CH inequality in this section.

 \subsection{The bound values for Bell-CH inequality}
 The bound values for the Bell-CHSH inequality  $\pm2\sqrt{2}$ are well
 known as Cirel'son bound \cite{C80}.
 The upper bound $(-1+\sqrt{2})/2$ of the Bell-CH inequality was proved by
  comparing the CH and CHSH inequalities \cite{Cabello}.
  The bound values for the Bell-CH inequality can also be simply found as
  follows.  The Bell-CH operator for a two-qubit system is 
 defined as \cite{CH,BW}
\begin{eqnarray}
&&{\cal B}_{CH}=  \xi_1(\theta_1)\otimes \xi_2(\theta_2)+ \xi_1(\theta_1)\otimes \xi_2(\theta^\prime_2)+\xi_1(\theta_1^\prime)\otimes \xi_2(\theta_2)\nonumber\\
&&~~~~~~- \xi_1(\theta_1^\prime)\otimes \xi_2(\theta_2^\prime)- \xi_1(\theta_1)\otimes \openone_2- \openone_1\otimes \xi_2(\theta_2) 
\label{ch},
\end{eqnarray}
where
\begin{eqnarray}
&&\xi(\theta)=|\theta\rangle\langle\theta|,\\
&&|\theta\rangle=\cos\theta|0\rangle+\sin\theta|1\rangle.
\end{eqnarray}
then the local theory imposes the inequality $-1\leq\langle{\cal
  B}_{CH}\rangle\leq0$.  Note here that we investigate a simple
$2\times2$ system without loss of generality. One can prove by direct
calculation
\begin{equation}
{\cal B}_{CH}^2=-{\cal B}_{CH}-\Delta,
\label{eq:sq}
\end{equation}
where
\begin{eqnarray}
&&\Delta=\langle\theta_1|\theta_1^\prime\rangle(|\theta_1\rangle\langle\theta_1^\prime|-|\theta_1^\prime\rangle\langle\theta_1|)_1\nonumber\\
&&~~~~~~~~~~~~~~~~~~~\otimes\langle\theta_2|\theta_2^\prime\rangle(|\theta_2\rangle\langle\theta_2^\prime|-|\theta_2^\prime\rangle\langle\theta_2|)_2.
\end{eqnarray}
Using $\langle{\cal B}_{CH}\rangle^2\leq\langle{\cal
  B}_{CH}^2\rangle$, the average of Eq.~(\ref{eq:sq}) becomes
\begin{equation}
\langle{\cal B}_{CH}\rangle^2+\langle{\cal B}_{CH}\rangle+\langle\Delta\rangle\leq0,
\end{equation}
and the Bell-CH function 
 $B_{CH}\equiv\langle{\cal B}_{CH}\rangle$ is
\begin{equation}
\frac{-1-\sqrt{1-4\langle\Delta\rangle}}{2}\leq B_{CH}
\leq\frac{-1+\sqrt{1-4\langle\Delta\rangle}}{2}.
\end{equation}
The maximal and minimal values of $\langle\Delta\rangle$ can be
obtained from the eigenvalues of $\Delta$ \cite{braun92}, which are
$\pm\sin[2(\theta_1-\theta_1^\prime)]\sin[2(\theta_2-\theta_2^\prime)]/4$.
The inequality $-1/4\leq\langle\Delta\rangle\leq1/4$ is then obtained.
Finally, the maximum and minimum of the Bell-CH function are found at
$\langle\Delta\rangle=-1/4$ as
\begin{equation}
\frac{-1-\sqrt{2}}{2}\leq 
B_{CH}\leq\frac{-1+\sqrt{2}}{2}
\end{equation}
in which the upper and lower bounds of the Bell-CH function
are given.
For example, the Bell-CH  function  for a single-photon entangled state
\begin{equation}
|\psi\rangle=\frac{1}{\sqrt{2}}(|0\rangle|1\rangle-|1\rangle|0\rangle)
\label{single}
\end{equation}
 is calculated to be
\begin{eqnarray}
&&B_{CH}=\frac{1}{4}\Big\{\cos[2(\theta_1^\prime-\theta_2^\prime)-\cos[2(\theta_1-\theta_2^\prime)\nonumber\\
&&~~~~~~~~~~~~~~-\cos[2(\theta_1^\prime-\theta_2)-\cos[2(\theta_1-\theta_2)]-2\Big\}.
\end{eqnarray}
This 
maximizes to 
 $(\sqrt{2}-1)/2\simeq0.21$ at $\theta_1=0$, $\theta_2=-3\pi/8$, $\theta_1^\prime=\pi/4$ and
$\theta_2^\prime=-5\pi/8$ \cite{eber93} and  minimizes to  
 $-(\sqrt{2}+1)/2\simeq-1.21$ at
$\theta_1=0$, $\theta_1^\prime=\pi/4$ and
$\theta_2=-\theta_2^\prime=\pi/8$.

\subsection{Bell-CH inequalities for continuous variables}

BW 
used  the $Q$ function for the test of the Bell-CH inequality violation 
 of the simple single-photon entangled state $(\ref{single})$
\cite{BW}. The $Q$ function for a two-mode state $\rho_{12}$ is defined as
\begin{equation}
Q_{12}(\alpha,\beta)=\frac{{}_2\langle\beta|{}_1\langle\alpha|\rho_{12}
|\alpha\rangle_1|\beta\rangle_2}{\pi^2},
\end{equation}
 where $|\alpha\rangle$
and $|\beta\rangle$ are coherent states.
The Bell-CH function in terms of $Q$ representation is 
\begin{eqnarray}
B_{CH-BW}=\langle \zeta_1(\alpha)\otimes
 \zeta_2(\beta)+\zeta_1(\alpha)\otimes \zeta_2(\beta^\prime)\nonumber\\
+\zeta_1(\alpha^\prime)\otimes \zeta_2(\beta)
-\zeta_1(\alpha^\prime)\otimes \zeta_2(\beta^\prime)\nonumber\\
-\zeta_1(\alpha)\otimes \openone_2
-\openone_1\otimes \zeta_2(\beta)
 \rangle  \nonumber\\ 
=\pi^2[Q_{12}(\alpha,\beta)+Q_{12}(\alpha,\beta^\prime)+Q_{12}(\alpha^\prime,\beta) \nonumber\\
-Q_{12}(\alpha^\prime,\beta^\prime)]-\pi[Q_1(\alpha)+Q_2(\beta)],
\label{ch-q}
\end{eqnarray}
where
 $Q_1(\alpha)$ and $Q_2(\beta)$  are the marginal $Q$ functions of modes
1 and 2, and 
 $\zeta(\alpha)=D(\alpha)|0\rangle\langle0|D^\dagger(\alpha)$. 
Eq.~(\ref{ch-q}) is a generalized version of the BW's formalism as BW
considered $\alpha=\beta=0$ \cite{BW}.
In this case the measurement
results are distinguished according to the presence of photons,
in other words, the dichotomic outcomes are no photon and the presence of photons.
  This
is more realistic because the parity of photon numbers is
difficult to measure with currently developed photodetectors.

\begin{figure}
\centerline{
{\scalebox{.6}
{\includegraphics{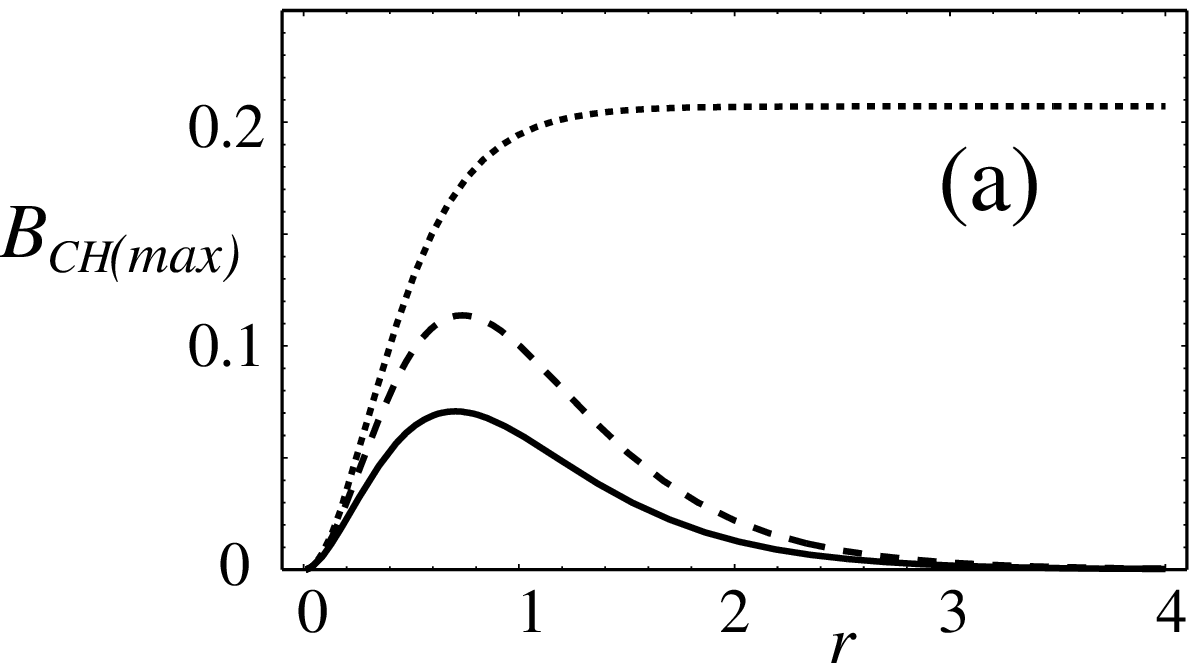}}}}\vspace{0.7cm}
\centerline{
{\scalebox{.6}
  {\includegraphics{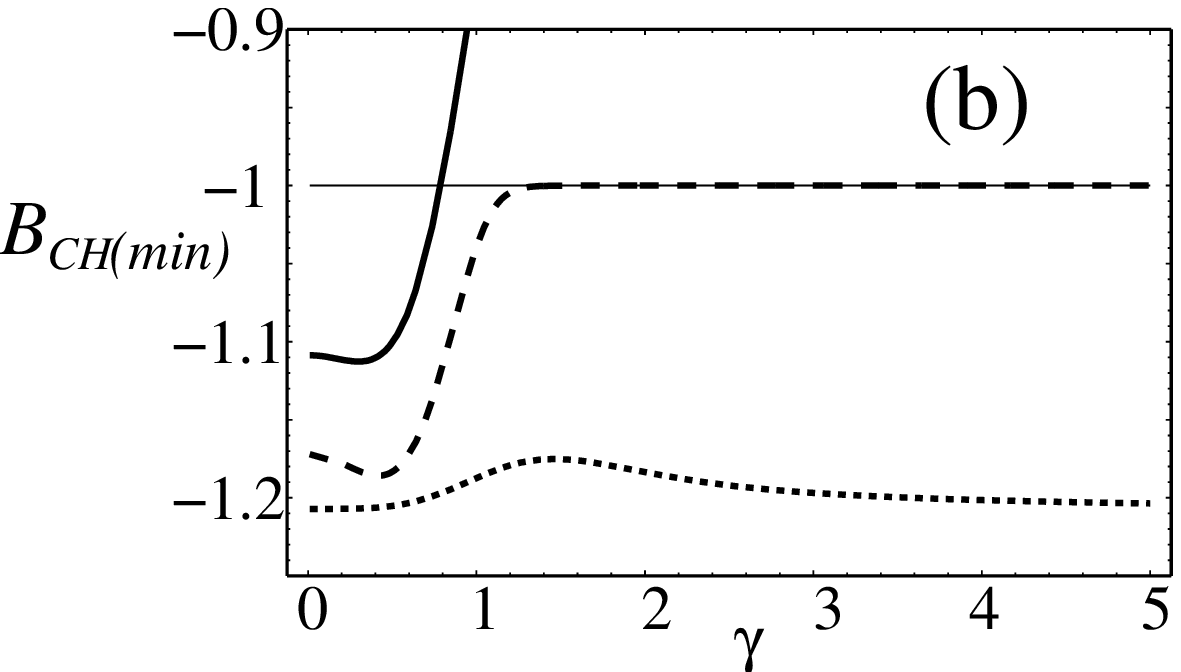}
}}}\vspace{0.2cm}
\caption{
  (a) The maximized Bell-CH function $B_{CH(max)}$ for a two-mode squeezed
  state is plotted against the degree of squeezing $r$ using the BW (solid line) and the
  generalized BW (dashed) formalisms. The maximized 
  function $B_{CH(max)}$ of the same state based upon parity measurement
  for the same state is given (dotted line).  (b) The minimized
  Bell-CH function $B_{CH(min)}$ for an entangled coherent state is plotted against its
  coherent amplitude $\gamma$ using the BW (solid line) and the
  generalized BW (dashed) formalisms. The
  minimized function $B_{CH(min)}$ based upon the parity measurement is plotted
   for the same state (dotted line).}
\label{fig3} 
\end{figure}

The $Q$ function for the two-mode squeezed state is \cite{BN} 
\begin{eqnarray}
Q_{TMSS}(\alpha,\beta)=\frac{1}{\pi^2\cosh^2 r}
\exp[-|\alpha|^2-|\beta|^2\nonumber\\
+\tanh r(\alpha\beta+\alpha^*\beta^*)]
\label{qnopa}
\end{eqnarray}
and the $Q$ function for the entangled coherent state 
\begin{eqnarray}
&&Q_{ECS}(\alpha,\beta)={\cal N}^2\Big\{
\exp[-|\alpha-\gamma|^2-|\beta+\gamma|^2]\nonumber\\
&&~~~~~~~~~~~~~~~~~~~~~~~~~~+\exp[-|\alpha+\gamma|^2-|\beta-\gamma|^2]\nonumber\\
&&~~-\exp[-(\alpha-\gamma)(\alpha^*+\gamma)-(\beta+\gamma)(\beta^*-\gamma)-4\gamma^2]\nonumber\\
&&~~-\exp[-(\alpha^*-\gamma)(\alpha+\gamma)-(\beta^*+\gamma)(\beta-\gamma)-4\gamma^2]
\Big\}.\nonumber\\
\label{qecs}
\end{eqnarray}
The marginal $Q$ function of each state can also be simply obtained from
Eqs.~(\ref{qnopa}) and (\ref{qecs}). 
One can
investigate the violation of the Bell-CH inequality 
for the two different states from Eqs.~(\ref{ch-q}), (\ref{qnopa}) and
(\ref{qecs}). The results are plotted in Figs.~\ref{fig3}(a) and (b).

For the two-mode squeezed state, the degree of the violation
of the Bell-CH inequality  
increases as generalizing the BW formalism. However, it increases up
to a peak and decreases as increasing the squeezing $r$, which is
shown in Fig.~\ref{fig3}(a).  
 The two-mode squeezed state is a separable pure state
when $r$ is zero, where no violation of the Bell's inequality is found.
 As $r$ increases, entanglement  
becomes to exist, which causes the violation of the Bell's inequality.
However, as $r$ increases, the average photon number increases and the
weight of $|0\rangle|0\rangle$ decreases as seen in Eq.~(\ref{eq:nopa}).
As the BW formalism of the Bell-CH violation is based on the nonlocality
of no photon and presence of photons, its violation diminishes when
$r$ is large.

The even and odd cat states become the no-photon and single-photon
number states respectively, {\it i.e.}
$|e\rangle\rightarrow|0\rangle$ and $|d\rangle\rightarrow|1\rangle$,
when $\gamma\rightarrow0$.  Therefore, the entangled coherent state
approaches to the single-photon entangled state (\ref{single}) in this
limit.  It can be simply shown that the degree of the Bell-CH
violation in the generalized BW formalism for the entangled coherent
state for $\gamma\rightarrow0$ ($B_{CH-BW}\simeq-1.17$) is the same as
the one for the single photon entangled state (\ref{single}).  It is
larger than the maximized value found by BW ($B_{CH-BW}\simeq-1.11$)
\cite{BW} which is also shown in Fig.~\ref{fig3}(b).  However, it does
not still reach the maximal violation $-(1+\sqrt{2})/2\simeq-1.21$,
which the single-photon entangled state (\ref{single}) shows with
perfect rotations.  It does not maximally violate the Bell-CH inequality
because of the imperfect rotations by the displacement operator used
in the BW formalism (\ref{ch-q}).  Note that the displacement operator
does not flip $|0\rangle$ to $|1\rangle$ and {\it vise versa} (see
Fig.~\ref{fig1}(b)).  As $\gamma$ becomes large, one can observe
qualitatively the same phenomenon as for the two-mode squeezed state.
The Bell violation approaches to zero as $\gamma\rightarrow\infty$
because of the decrease of the weight of the term
$|0\rangle|0\rangle$.

Instead of the measurement of the presence of photons, the parity
measurement can be used with the unitary rotation $U(\theta)$ to
investigate the Bell-CH inequality.  The Bell-CH function is defined
as
\begin{eqnarray}
&&B_{CH}^{(\Pi)}=\langle \chi_1(\theta_1)\otimes
 \chi_2(\theta_2)+\chi_1(\theta_1)\otimes \chi_2(\theta_2^\prime)\nonumber\\
&&~~~~~~~~~~~~~~~+\chi_1(\theta_1^\prime)\otimes \chi_2(\theta_2)
-\chi_1(\theta_1^\prime)\otimes \chi_2(\theta_2^\prime)\nonumber\\
&&~~~~~~~~~~~~~~~-\chi_1(\theta_1)\otimes \openone_2
-\openone_1\otimes \chi_2(\theta_2)
 \rangle,  \\
&&\chi(\theta)=\sum_{n=0}^\infty U(\theta)|2n\rangle\langle 2n|U^\dagger(\theta).
\end{eqnarray}
For the two-mode squeezed state,
\begin{eqnarray}
\langle\chi_1(\theta_1)\otimes
\chi_2(\theta_2)\rangle=\sin\theta_1\cos\theta_1\sin\theta_2\cos\theta_2
\tanh 2r,\\
\langle\chi_1(\theta_1)\otimes
\openone\rangle=\frac{(\cos^2\theta_1\cosh^2 r+\sin^2\theta_1^2\sinh^2
  r)}{\cosh 2r},
\end{eqnarray}
and for the entangled coherent state,
\begin{eqnarray}
&&\langle\chi_1(\theta_1)\otimes
\chi_2(\theta_2)\rangle=
\frac{1}{2}(\sin^2\theta_1\cos^2\theta_2+\sin^2\theta_1\cos^2\theta_2)\nonumber\\
&&~~~~~~~~~~~~~~~~~~~~~~~~~-K(\gamma)\sin\theta_1\cos\theta_1\sin\theta_2\cos\theta_2,\\
&&\langle\chi_1(\theta_1)\otimes
\openone\rangle=\frac{1}{2}(\cos^2\theta_1+\sin^2\theta_1),
\end{eqnarray}
from which the Bell-CH function $B_{CH}^{(\Pi)}$ can be obtained.
In both cases, we find that the Bell-CH function approaches to the
 maximal violation $B_{CH} 
^{(\Pi)}\rightarrow-(1\pm\sqrt{2})/2$.  For the two-mode squeezed
state, $B_{CH}^{(\Pi)}$ reaches  the maximal violation 
for $r\rightarrow\infty$ as shown in Fig.~\ref{fig3}(a).
The upper bound is found at 
$\theta_1=0$, $\theta_2=-3\pi/8$,
$\theta_1^\prime=\pi/4$ and $\theta_2^\prime=-5\pi/8$,
and the lower bound at
$\theta_1=0$, $\theta_2=-\theta_2^\prime=\pi/8$ and
$\theta_1^\prime=\pi/4$.
As shown in Fig.~\ref{fig3}(b), for the entangled coherent state,
$B_{CH}^{(\Pi)}$ reaches  the
maximal violation 
for $\gamma\rightarrow0$ and $\gamma\rightarrow\infty$
 at the same angles.

\section{remarks}
We have studied the violation of Bell's inequalities using various
formalisms.  We have been able to discuss the link between the
discussions for the quantum nonlocality of finite and infinite dimensional
systems.  The pseudospin operator \cite{Chen01} can be understood as
the limiting case of Gisin-Peres observable \cite{GP}.  The BW
formalism \cite{BW} can be generalized to obtain a larger Bell
violation \cite{Derek}. However, the original EPR state cannot
maximally violate Bell's inequality even in the the generalized
version of the BW formalism.  We discussed the reason compared with
the case of the entangled coherent state which shows maximal violation
of Bell's inequality in the generalized BW formalism.  Our result is
in agreement with the recent study of nonlocality of a two-mode
squeezed state in absorbing optical fibers \cite{Filip}.  In
\cite{Filip}, the authors found that nonlocality of the two-mode
squeezed state is more robust against a dissipative environment in
pseudospin approach than in the previous study \cite{JLK00} based on
the BW formalism.  It was shown that the dichotomic measurement for
the presence of photons is not so effective in finding the nonlocality
of two-mode squeezed states and entangled coherent states.

However, it must be pointed out that the nonlocality based
on the Wigner and $Q$ functions is extremely useful because we know
the measurement of $W$ and $Q$ functions is experimentally possible while
the implementation of other operations which we have
discussed here have difficulties in their experimental realization.

{\it Note added in proof.}  After submitting our paper,
we have found that Banaszek {\it et al.} studied the Bell-CH inequality for a
single-photon entangled state and a two-mode squeezed state in terms
of $Q$ representation \cite{r25}. They took imperfect detection efficiency into consideration.
Recently, a paper generalizing the work
of Chen {\it et al.} where different qubit states are assigned to a
continuous variable system has also appeared \cite{r26}.

\acknowledgements
HJ thanks J. Lee for discussions. 
This work has been supported by the UK Engineering and Physical
Sciences Research Council (EPSRC) (GR/R 33304)
and the Korean Ministry of Science and Technology through the Creative
Research Initiative Program under Contract No. M1-0116-00-0008. 
{\v C}B is supported by the Austrian FWF Project No. F1506 and by the
European Commission, Contract No. ERBFMRXCT96-0087.

\end{document}